\documentclass[aps,prl,twocolumn,amsmath,amssymb,superscriptaddress,10pt,nobibnotes]{revtex4-2}  

\usepackage{ulem}

\usepackage{graphicx}
\usepackage{dcolumn}
\usepackage{bm}
\usepackage{subfig}

\begin{document}

\title{Shaping an evanescent focus of light for high spatial resolution optogenetic activations in live cells}

\author{Marc GROSJEAN}
\affiliation{Univ. Grenoble Alpes, CNRS, LIPhy, F-38000 Grenoble, France}
\author{Alexei GRICHINE}
\affiliation{MicroCell platform, Institute for Advanced Biosciences (IAB), Centre de Recherche UGA / Inserm U 1209 / CNRS UMR 5309, France}
\author{ Mylene PEZET}
\affiliation{MicroCell platform, Institute for Advanced Biosciences (IAB), Centre de Recherche UGA / Inserm U 1209 / CNRS UMR 5309, France}
\author{Olivier DESTAING}
\affiliation{MicroCell platform, Institute for Advanced Biosciences (IAB), Centre de Recherche UGA / Inserm U 1209 / CNRS UMR 5309, France} 
\affiliation{Team "Epigenetic, Immunity, Metabolism, cell signaling and cancer", Institute for Advanced Biosciences (IAB), Centre de Recherche UGA / Inserm U 1209 / CNRS UMR 5309, France}
\author{Antoine DELON}
\affiliation{Univ. Grenoble Alpes, CNRS, LIPhy, F-38000 Grenoble, France}
\affiliation{MicroCell platform, Institute for Advanced Biosciences (IAB), Centre de Recherche UGA / Inserm U 1209 / CNRS UMR 5309, France} 
\author{Ir\`ene WANG}
\affiliation{Univ. Grenoble Alpes, CNRS, LIPhy, F-38000 Grenoble, France}

\email{irene.wang@univ-grenoble-alpes.fr}

\begin{abstract} 

Confining light illumination in the three dimensions of space is a challenge for various applications. Among these, optogenetic methods developed for live experiments in cell biology would benefit from such a localized illumination as it would improve the spatial resolution of diffusive photosensitive proteins leading to spatially constrained biological responses in specific subcellular organelles. Here, we describe a method to create and move a focused evanescent spot, at the interface between a glass substrate and an aqueous sample, across the field of view of a high numerical aperture microscope objective, using a digital micro-mirror device (DMD). We show that, after correcting the optical aberrations, light is confined within a spot of sub-micron lateral size and $\sim$100~nm axial depth above the coverslip, resulting in a volume of illumination drastically smaller than the one generated by a standard propagative focus. This evanescent focus is sufficient to induce a more intense and localized recruitment compared to a propagative focus on the optogenetic system CRY2-CIBN, improving the resolution of its pattern of activation.

\end{abstract}

\maketitle
\section*{Introduction}

Evanescent waves are of interest in many applications, since they make it possible to confine light in a layer, typically of a few hundred nanometers, near an interface. For example, they can be used in spectroscopy to measure the absorption of interfacial layers~\cite{Kazarian2013}, or to print ultra-thin structures~\cite{You2019}. In fluorescence microscopy, the total internal reflection fluorescence (TIRF) technique takes advantage of evanescent waves to reduce out-of-focus background and significantly improves signal-to-noise ratio in images of interfaces of living samples ~\cite{ Axelrod2013,MARTINFERNANDEZ2013}. TIRF is widely used to observe cell membranes \cite{Axelrod1981,Mattheyses2010}, as well as for single molecule localization-based super-resolution imaging~\cite{MOERNER2012}. Compared to epifluorescence and confocal imaging, TIRF microscopy leads to reduced photobleaching and photoxicity, since fluorophores are only excited in the vicinity of the sample interface instead of the whole sample thickness, allowing to observe living cells for longer times.

During the last decade, optogenetics has emerged as an essential tool to analyse dynamic processes in neuroscience and cell biology~\cite{Goglia2019}. Indeed, optogenetics combines the high specificity of genetics with photosensitive elements allowing to use light in a given spectral range to control protein activity in real time, unveiling the spatiotemporal aspects of intracellular processes. At the cellular and subcellular levels, optogenetics has been massively used for temporal control while spatial control has been poorly investigated. Indeed, probing the cell's  spatial response requires generating well-defined spatial activation patterns. This could be achieved with few micrometers resolution in reversible optogenetic systems by projecting two light patterns of different wavelengths to activate a pattern while inactivating regions outside of it~\cite{Levskaya2009}. However, for most optogenetic systems, achieving micron-scale local activation is not straightforward. This is the case of CRY2/CIBN~\cite{Kennedy2010}, a widely-used system to probe cell signalling.  When CIBN is tethered to the plasma membrane light illumination induces a hetero-oligomerization of CRY2 and CIBN causing the translocation of the effector to the membrane, which triggers a biological response. The photosensitive protein CRY2 is also known to oligomerize upon light activation~\cite{Bugaj2013}, a phenomenon that was used to control the spatial distribution of an effector fused to it through a game of affinities between monomers and oligomers~\cite{Kerjouan2021}. In any case, the spatial resolution of activated regions is limited by the diffusion of CRY2 complexes in the cytosol (3D) and at the membrane (2D)~\cite{Valon2015}, so that the smallest region that can be activated is around 5-10~\textmu m. Evanescent light has been used for photoactivation in a TIRF configuration~\cite{Kerjouan2021}. In this case, although cytosolic diffusion can be neglected (as only proteins close to the membrane are activated that can rapidly bind  to their membrane partners), standard TIRF illuminates a wide region of the sample and does not allow spatial patterning. 

In the present work, we aim at generating an evanescent spot of light that is confined in the three dimensions of space at the surface of a glass substrate. By scanning such a spot, one would be able to produce an arbitrary evanescent pattern, in order to photoactivate proteins at the cellular membrane with an optimal spatial resolution. In the beginnings of TIRF microscopy, Axelrod and Stout~\cite{Stout1989} showed the possibility to create such an evanescent spot, with a 1.5~\textmu m radius, by sending a laser beam in a peripheral ring in the back focal plane  (BFP) of a high aperture objective, using an opaque mask. However, the spot created in their configuration is  located on the optical axis and cannot be moved, which limits the use of this configuration in biological applications.

Here, we propose to generate an evanescent spot that can be moved on a 100~\textmu s timescale, using a digital micro-mirror device (DMD). DMDs exhibit a large number of pixels together with fast switching rates, so that they have found various applications in microscopy, including structured illumination \cite{Chakrova2015}, scanning microscopy~\cite{Dussaux2018} or wavefront shaping in complex media \cite{Conkey2012}. They have also been used for TIRF microscopy, but only to avoid shadows by scanning the illumination angle~\cite{Zong2014}. In our application, we show that the DMD is a versatile solution to (i) mask the central area of the BFP corresponding to propagating waves, (ii) display the evanescent spot at chosen positions, and (iii) correct the aberrations of the microscope setup. The resulting evanescent spot has been characterised in terms of lateral width and axial penetration depth. Finally, preliminary photoactivation experiments on living cells expressing the CRY2-CIBN system indicate the potential of an evanescent spot not only to reduce the size of the distribution of recruited CRY2, but also to enhance the level of recruitment at the position of interest.

\section*{Method}

\begin{figure*}
    \centering
    \includegraphics[width=1\linewidth]{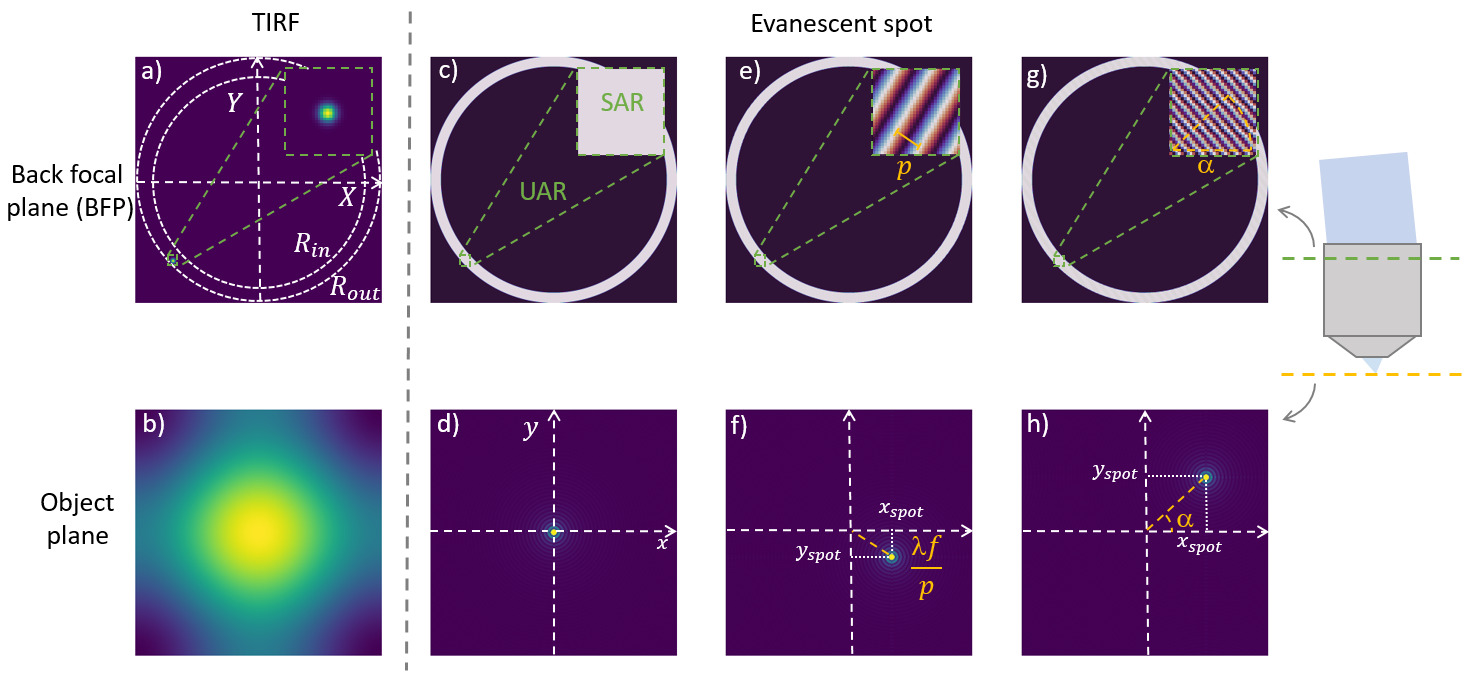}
    \caption{\textit{(a, b)} Classical standard TIRF configuration. The light beam is focused in the periphery of the BFP in order to reach the interface between the two medium at $\theta > \theta_c$, resulting in an evanescent but wide illumination in the object plane. \textit{(c,d)} Uniform phase map in the BFP and the corresponding amplitude distribution in the object plane. \textit{(e,g)} Phase maps in the case of tilted wavefronts. The light intensity (not represented) is equal to 1 in the SAR and 0 elsewhere. \textit{(f,h)} Corresponding amplitude distribution in the object plane for the tilted wavefronts. Note that, as $p$ decreases, the spot moves away from the optical axis in the direction of angle $\alpha$.} 
    \label{simulation}
\end{figure*}

In TIRF microscopy, evanescent waves are created by using the total internal reflection phenomenon. When a plane wave, propagating in a medium with a refractive index $n_1$, reaches an interface with another medium with a lower index $n_2$, at an incidence angle $\theta$ larger than the critical angle, $\theta_c = \arcsin(\frac{n_2}{n_1})$, the propagating wave is totally reflected in the first medium while an evanescent wave is emitted in the second medium. In TIRF, this critical angle is reached by focusing the light in the periphery of the BFP of an immersion objective with a high numerical aperture ($\rm{NA}>1.33$). In the BFP of such an objective, we can define two different regions. Let us note $(X,Y)$ the coordinates in the BFP, with $R = \sqrt{X^2 + Y^2}$, and $R_{out}$ the radius of the objective pupil (assumed to be located at the BFP). The first region is a peripheral ring $R_{in} < R < R_{out}$, with $R_{in} = \frac{n_2}{NA}\times R_{out}$, corresponding to light that reaches the interface of the two medium with angles $\theta$ > $\theta_c$ ; it will be named the \textit{supercritical angle region} (SAR). The second region in the BFP is the inner disk $R <R_{\rm{in}}$ of this peripheral ring, corresponding to incident angles, $\theta<\theta_c $ and will be named the \textit{under-critical angle region} (UAR), as shown on Fig.~\ref{simulation}.

In standard TIRF, to pass through the fine peripheral region, the excitation laser is focused in the BFP, so that it illuminates a very large area in the object plane, typically hundreds of \textmu m, which is useful for imaging purposes. However, if rather than focusing light onto a point in the BFP, the SAR is entirely illuminated with a planar wavefront, the radial extension of the beam in the object plane will be drastically smaller with a typical size limited by diffraction, as illustrated on Fig.~\ref{simulation}. This configuration has already been described in Ref.~\cite{Stout1989} with an immobile evanescent spot in the object plane, but the usefulness of such an immobile single spot is limited. To move the spot, the BFP must be illuminated with a tilted planar wave, as it can be seen in Fig.~\ref{simulation}.

The tilted plane wave, that focuses at a position $(x_{\rm{spot}}, y_{\rm{spot}})$  in the object plane, can be written in the BFP as :
\begin{equation}
    \psi(X, Y) = A(X,Y) e^{i(k_XX+k_YY)} 
\end{equation}
with $k_X = \frac{2\pi}{\lambda} \frac{x_{\rm{spot}}}{f} $, $ k_Y = \frac{2\pi}{\lambda} \frac{y_{\rm{spot}}}{f}$ and $f$, the focal length of the objective. If the amplitude is uniform in the SAR and nil outside: 
\begin{equation}
A_{\rm{SAR}}(X,Y) = \left\{\begin{array}{lll}
         1 & \rm{if} & R_{\rm{in}} < R < R_{\rm{out}} \\
         0 & \rm{elsewhere} & 
    \end{array}
\right. 
\label{ASAR}
\end{equation}
the focus in the sample plane is evanescent. In the following, we'll refer to it as "evanescent spot" or "evanescent focus". On the other hand, if the amplitude is uniform in the UAR and nil outside: 
\begin{equation}
A_{\rm{UAR}}(X,Y) = \left\{\begin{array}{lll}
         1 & \rm{if} &  R < R_{\rm{in}} \\
         0 & \rm{elsewhere} & 
    \end{array}
\right. 
\label{AUAR}
\end{equation}
the focus in the sample plane is propagative. In the following, we'll refer to it as "propagative spot" or "propagative focus". The propagative focus, which would be observed in a confocal microscope with the same objective, is used throughout this work as a comparison with the evanescent focus.

When generating a tilted wavefront, the phase is wrapped and looks like a 2D saw pattern of period $p = \frac{2\pi}{\sqrt{{k_X}^2+{k_Y}^2}}$
along a direction given by the angle, $\alpha = \arctan(\frac{k_Y}{k_X})$. As shown in Fig.~\ref{simulation}~(e-h), when $p$ decreases, the spot moves away from the optical axis in the direction given by $\alpha$.

In order to create the tilted wave, we use a DMD. The DMD is a binary amplitude controller. It consists of an array of square micro-mirrors of dimension, $d$=7.6~\textmu m, and is placed in a plane conjugated to the BFP. Each micro-mirror can either reflect the light in the direction of the BFP (ON state) or elsewhere (OFF state). To create a tilted wave with micro-mirrors, we display a binary fringe pattern of period $p$ and angle $\alpha$, according to the following formula:
\begin{equation}
\rm{S_{DMD}}(X',Y') = \left\{
    \begin{array}{lllll}
         ON & \rm{if} & \cos(k_X \frac{X'}{\gamma_{\rm{DMD}}}+ k_Y \frac{Y'}{\gamma_{\rm{DMD}}} ) > 0 \\
         OFF & \rm{elsewhere}
    \end{array}
\right.
\label{equation1}
\end{equation}
where $(X',Y')$ are the coordinates in the DMD plane, which are magnified by a factor $\gamma_{\rm{DMD}} = f_{L4}/f_{L2}$ compared to those in the BFP ($f_{L2}$ and $f_{L4}$ are the focal lengths of the respective lenses depicted in Fig.~\ref{setup}). Note that this modulation is displayed only in the SAR for an evanescent focus, and only in the UAR for a propagative focus, by taking the product between $\rm{S_{DMD}}$ and the amplitude function given in Eq.~\ref{ASAR} or \ref{AUAR} expressed in the DMD plane. The resulting far-field diffraction is given by the Fourier series representation of this pattern :
\begin{equation}
\rm{S_{DMD}}(X',Y') = \sum_{m=-\infty}^{\infty} \frac{\sin(\pi m)}{\pi m}e^{i m(k_X \frac{X'}{
\gamma_{\rm{DMD}}}+ k_Y \frac{Y'}{\gamma_{\rm{DMD}}})} 
\end{equation}
The first diffracted order ($m=1$) has exactly the desired phase. By filtering the other orders in the Fourier space of the DMD, we obtain the desired tilted wavefront in the BFP. Now, simply by changing the parameters $p$ and $\alpha$, we can illuminate different positions of the object plane with an evanescent spot.

Note that galvanometric mirrors could also have been used to tilt the wavefront and move the focus. However, this would require adding several lenses and an annular mask, leading to a more complex setup. In addition, such a solution does not allow the correction of possible optical aberrations.

\section*{Setup}

Our setup is based on an inverted microscope stand (Olympus IX73) equipped with a sCMOS camera (ORCA-Flash4.0, Hamamatsu). A custom light path has been added that allows to switch between standard TIRF, epifluorescence (EPI) and evanescent or propagative focus configurations.

Two fiber-coupled laser sources are used: the first one at $\lambda_{\rm{observation}} = 561$~nm (Oxxius LCX-561), and the second at $\lambda_{\rm{excitation}} = 488$~nm (Integrated Optics 0488L-21A). The fiber tip of the 561~nm laser is conjugated with the objective BFP, Olympus APO N 60x/1.49 oil, in order to perform standard TIRF imaging.

\begin{figure*}
    \centering
    \includegraphics[width=0.8\linewidth]{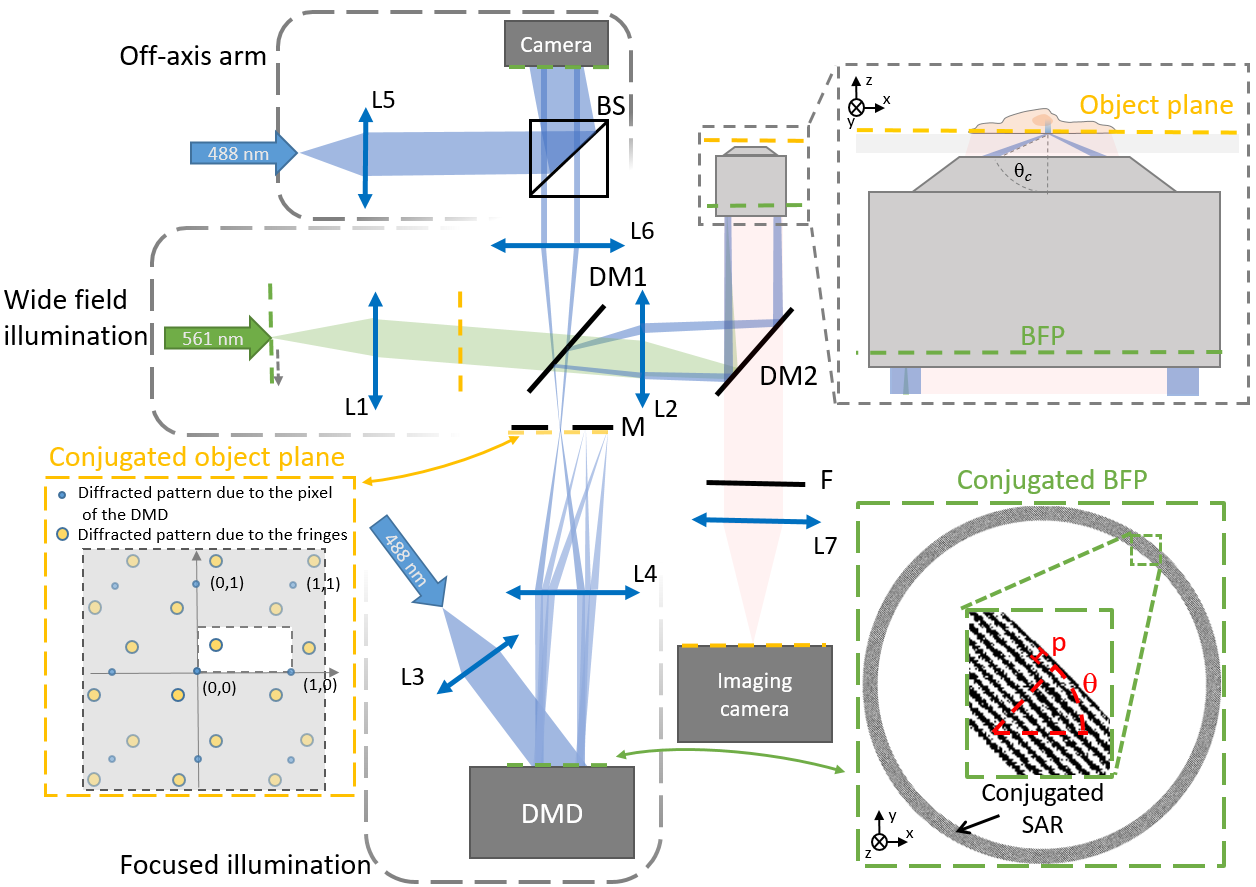}
    \caption{Experimental setup. A 561~nm laser is  used for wide-field illumination of the sample, either in TIRF or epi. In a second path, a 488~nm stimulation laser is reflected by a DMD and sent towards the microscope objective. The DMD, conjugated with the objective pupil plane, displays a binary pattern in the SAR region, thanks to its micro-mirrors of dimension $d = 7.6$~\textmu m, to produce an evanescent focus at any chosen position in the sample plane. The light leaking through DM1 is used to perform off-axis holography in order to measure the aberrations. (Lenses: L1 ($f=60$ mm), L2 ($f=150$ mm), L3 ($f=60$ mm), L4 ($f=200$ mm), L5 ($f=60$mm), L6 ($f=120$mm) - Dichroic mirror: DM - Filter: F - Mask: M - Beam splitter: BS.) }
    \label{setup}
\end{figure*}

In order to create the focus, the 488~nm beam is collimated and sent onto a DMD (Vialux DLP9000) which is in a conjugate plane of the BFP of the objective (Fig.~\ref{setup}). For the evanescent focus, only the micro-mirrors of the DMD conjugated with the SAR reflect the light in the direction of the BFP. Those micro-mirrors consist of a ring of approximately $3.10^5$ mirrors which corresponds to $\simeq16\%$ of the whole pupil surface. In practice, to ensure that we have only evanescent light, the width of the ring is slightly reduced, so that the inner radius corresponds to a numerical aperture of 1.36, a bit larger than the 1.33 required for an aqueous sample. Relay lenses in a 4f-configuration conjugate the DMD to the BFP, with a magnification of $\gamma_{\rm{DMD}} = \frac{4}{3}$ to match the size of the DMD to that of the pupil.

A first dichroic mirror DM1 (Semrock Di03-R488-t3) is used to combine the two lasers. The fluorescence emission is filtered by another dichroic mirror DM2 (Semrock Di03-R488/561-t3) and two filters F (Semrock 500/LP and Chroma ZET561NF) before reaching the imaging camera (Hamamatsu sCMOS ORCA-Flash4.0 V3).

In order to correct the aberrations induced by the DMD, we measure the wavefront of the beam after the DMD by off-axis interferometry. A fraction of the beam, transmitted by the first dichroic, interferes with a reference beam in a plane equivalent to the objective BFP. The resulting fringes are measured by a CMOS camera (Basler acA2440-35um). For practical reasons, the camera is not placed after the second lens of the 4f system leading to the BFP, but after an equivalent lens.

As explained in the previous section, the DMD is used to display a binary fringe pattern in the ring corresponding to the SAR region. The diffraction figure associated to those fringes produces an infinite number of spots in the sample plane, but fortunately, in practice, the intensity of the orders beyond the first one is negligible. Therefore, we only need to filter out the spots corresponding to 0 and the -1 orders. Moreover, the DMD micro-mirrors are themselves arranged in a regular array, which induces a square diffracted pattern (on a larger scale than the one induced by the displayed fringes). The final pattern of spots in the sample plane is a convolution of this square grid with the diffraction pattern created by the fringes. Hence, a large number of unwanted replica spots are potentially present. One spot can be unambiguously generated in a region delimited between the $0^{th}$ order, $\frac{\lambda f}{2d}$ in one direction and $\frac{\lambda f}{d}$ in the orthogonal direction, corresponding to a rectangle of $\simeq200$~\textmu m$\times 100$~\textmu m in the sample plane. We limit the area in which we can focus the beam, to this region by placing a rectangular mask (in a plane conjugated with the object plane) which transmits half the area delimited by the orders (0,0), (0,1), (1,0) and (1,1) diffracted by the micro-mirrors array, as shown in Fig.~\ref{setup}.

\section*{Correction of the aberrations}

\begin{figure*}[ht!]
    \centering

    \includegraphics[width=0.7\linewidth]{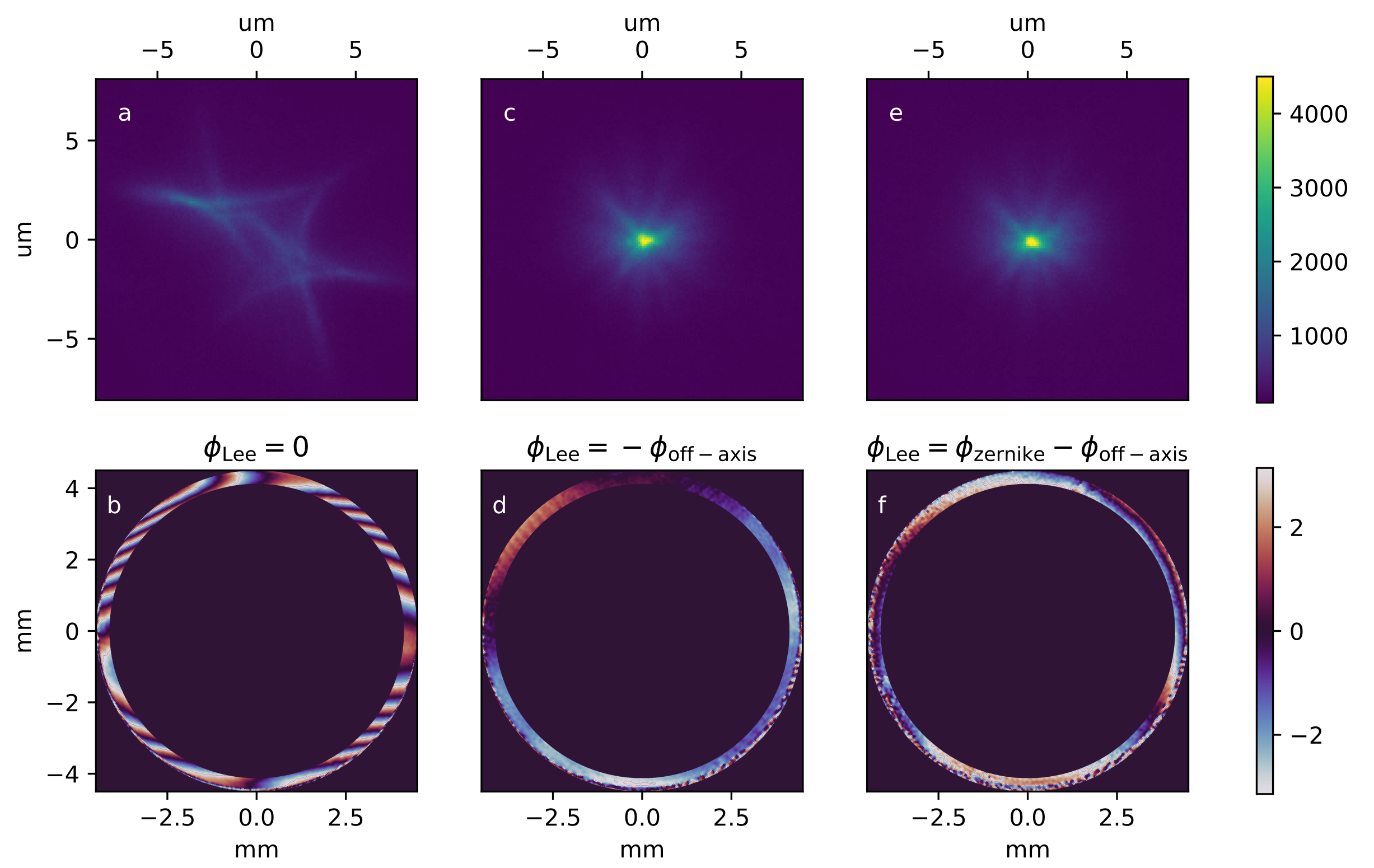}
    \caption{Aberration correction. Without phase correction ($\phi_{\rm{Lee}} = 0$), the evanescent spot in the object plane is highly distorted \textit{(a)}. The phasemap measured on the off-axis arm in this case is noted $\phi_{\rm{off-axis}}$ \textit{(b)}. When the correction $\phi_{\rm{Lee}} = -\phi_{\rm{off-axis}}$ is applied, the spot shape is greatly improved \textit{(c)} and the phase measured by the off-axis technique is almost flat \textit{(d)}. After a second correction step that consists in optimizing each Zernike mode, the spot is slightly more intense \textit{(e)}, while the phase measured on the off-axis arm shows a small deformation . }
    \label{aberration}
\end{figure*}

As any real optical system, our setup is not free from optical geometric aberrations and their correction is crucial in order to attain the highest possible spatial resolution. This is particularly important as the DMD is not flat at the wavelength scale, so that the resulting evanescent spot is strongly blurred as shown on Fig.~\ref{aberration}(a)

As presented above, in order to create a tilted wavefront in the BFP of the objective, we display a binary amplitude fringe pattern of period $p$. We can take advantage of this fringe pattern to implement the binary holography technique introduced by Lee~\cite{Lee1978}. This technique makes it possible to create a phase modulation by using a binary fringe pattern after selection of the first diffracted order.

The technique consists simply in adding the desired phase modulation $\phi_{\rm{Lee}}$ to Eq.~\ref{equation1} such that :

\begin{widetext}
\begin{equation}
    \rm{S_{DMD}}(X',Y') = \left\{
    \begin{array}{lll}
         \rm{ON} & \rm{if} & \cos[k_X \frac{X'}{\gamma_{\rm{DMD}}} + k_Y \frac{Y'}{\gamma_{\rm{DMD}}} + \phi_{\rm{Lee}}(X',Y')] > 0 \\
         \rm{OFF} & \rm{elsewhere}
    \end{array}
\right.
\end{equation}
\end{widetext}
The Fourier series representation of this pattern is:
\begin{widetext}
\begin{equation}
\rm{S_{DMD}}(X',Y') = \sum_{m=-\infty}^{\infty} \frac{\sin(\pi m)}{\pi m}e^{i m(k_X \frac{X'}{\gamma_{\rm{DMD}}} + k_Y \frac{Y'}{\gamma_{\rm{DMD}}}  + \phi_{\rm{Lee}}(X',Y') )} 
\end{equation}
\end{widetext}

Once again, if we keep only the first diffracted order, we end up with a complex field proportional to $e^{i(k_X \frac{X'}{\gamma_{\rm{DMD}}} + k_Y \frac{Y'}{\gamma_{\rm{DMD}}} + \phi_{\rm{Lee}}(X',Y'))}$. In this way, the phase modulation $\phi_{\rm{Lee}}(X',Y')$ can be applied onto the wavefront incident on the BFP.

This method can be used to correct the optical aberrations. Here and in the following, a homogeneous solution (Sulforhodamine B in water) is used as a test sample: the fluorescence excited by an evanescent spot close to the coverslip and acquired by the imaging camera, will be named 'image of the evanescent spot'; in the same way, we will write 'image of the propagative spot' to refer to the fluorescence induced by a propagative spot. In the following, the phase distortion induced by all the optical setup up to the evanescent spot is called $\phi_{\rm{aberration}}$ (as expressed in the objective pupil plane). To compensate for it, we need to apply a phase map $\phi_{\rm{Lee}}$ which satisfies $\phi_{\rm{Lee}}(X',Y') = - \phi_{\rm{aberration}}(X',Y')$. 

In order to determine $\phi_{\rm{aberration}}(X',Y')$ as accurately as possible, we followed a two-step procedure. The first step consists in a direct measurement of the wavefront using off-axis holography~\cite{Cuche2000}. The wavefront is measured after the reflection on the DMD, as the DMD was found to be the predominant source of aberrations. The measured phase, named $\phi_{\rm{off-axis}}$, Fig.~\ref{aberration}(b), can then be corrected by displaying $\phi_{\rm{Lee}}=-\phi_{\rm{off-axis}}$ on the DMD. On Fig.~\ref{aberration}(c), we show the evanescent spot after this first step.

To get the best possible correction, we performed a second step, in which the wavefront is iteratively optimized while monitoring the image of the evanescent spot. This
procedure is similar to well-known sensorless approaches for aberration correction in confocal or two-photon microscopy~\cite{Debarre:07}. We used as optimization metric the maximum intensity in the image and sequentially optimized each aberration mode (the modes being described as Zernike polynomials truncated on the annular SAR region). Note that standard circular Zernike polynomials are not an orthogonal mode basis in our case, since we use an annular pupil. A modified basis (annular Zernike  polynomials) has been proposed previously to provide orthogonal modes on such a pupil \cite{Mahajan1981}. However, in practice, we have not observed a more efficient or more accurate correction when using those modified modes. That is why the optimizations have been performed with the standard Zernike polynomials. The phase corrected in this second step may be written as:  
\begin{equation}
    \phi_{\rm{Zernike}} = \sum^{15}_{j=4} a_j Z_j 
\end{equation}
where $Z_j$ are truncated Zernike polynomials, numbered in the way described by Mahajan~\cite{Mahajan1994}. We used 12 modes for the optimization (from mode 4 to 15, since the first three modes : piston, tip and tilt are excluded).

To determine the amplitudes $a_j$, we followed the algorithm: for each mode $j$ in turn, the spot maximum intensity $I$ is measured for different known amplitudes (bias) of this mode applied by the DMD. We perform 20 measurements of $I$ for each mode. The amplitude corresponding to the maximum of $I$, which provides the optimal correction for the considered mode ($a_j$), is obtained from a parabolic fit. This correction is immediately applied on mode $j$, before we proceed to the next mode.

One can iterate several times over all the modes to account for large aberration amplitudes and also to make the process more robust to the presence of crosstalk between modes. In our case, we observed that, after a second loop, the value of $I$ does not continue to increase. Hence, we generally limit the procedure to two iterations.

The final aberration correction applied on the DMD is then a superposition of the phase maps determined from the two steps: $\phi_{\rm{Lee}} =  - \phi_{\rm{off-axis}} + \phi_{\rm{Zernike}}$. 
As shown on Fig.~\ref{aberration}(a), initially, the spot quality is largely distorted but thanks to this correction we retrieve a spot with a significantly better shape, Fig.~\ref{aberration}(e). The spot quality is drastically improved by the first step of the correction, while the second step leads to a small increase of the intensity at the spot centre.

We observed a similar spot quality after performing aberration correction, whatever the position of the evanescent spot within the field-of-view. However, the aberrations may vary depending on the spot position, which limits the validity of a given wavefront correction. Since the distortion due to the DMD (the major source of aberrations in our case) does not change in the field-of-view, the shape of the focus was found to remain equivalent  within a disk of 100~\textmu m in diameter around the correction point. This is in agreement with what is commonly observed in microscopes.

\section*{Characterisation of the evanescent focus spatial extension}

\subsection*{Lateral size}
\label{lateral_size}
To measure the lateral size of the evanescent spot, as previously, we used a fluorescent  solution. Two configurations have been compared: light was sent either onto the annular SAR of the objective BFP leading to an evanescent focus in the sample, or onto a disk corresponding to the UAR of the objective BFP resulting in a propagative focus in the sample.

\begin{figure*}[ht!]
    \centering
    \includegraphics[width=0.8\textwidth]{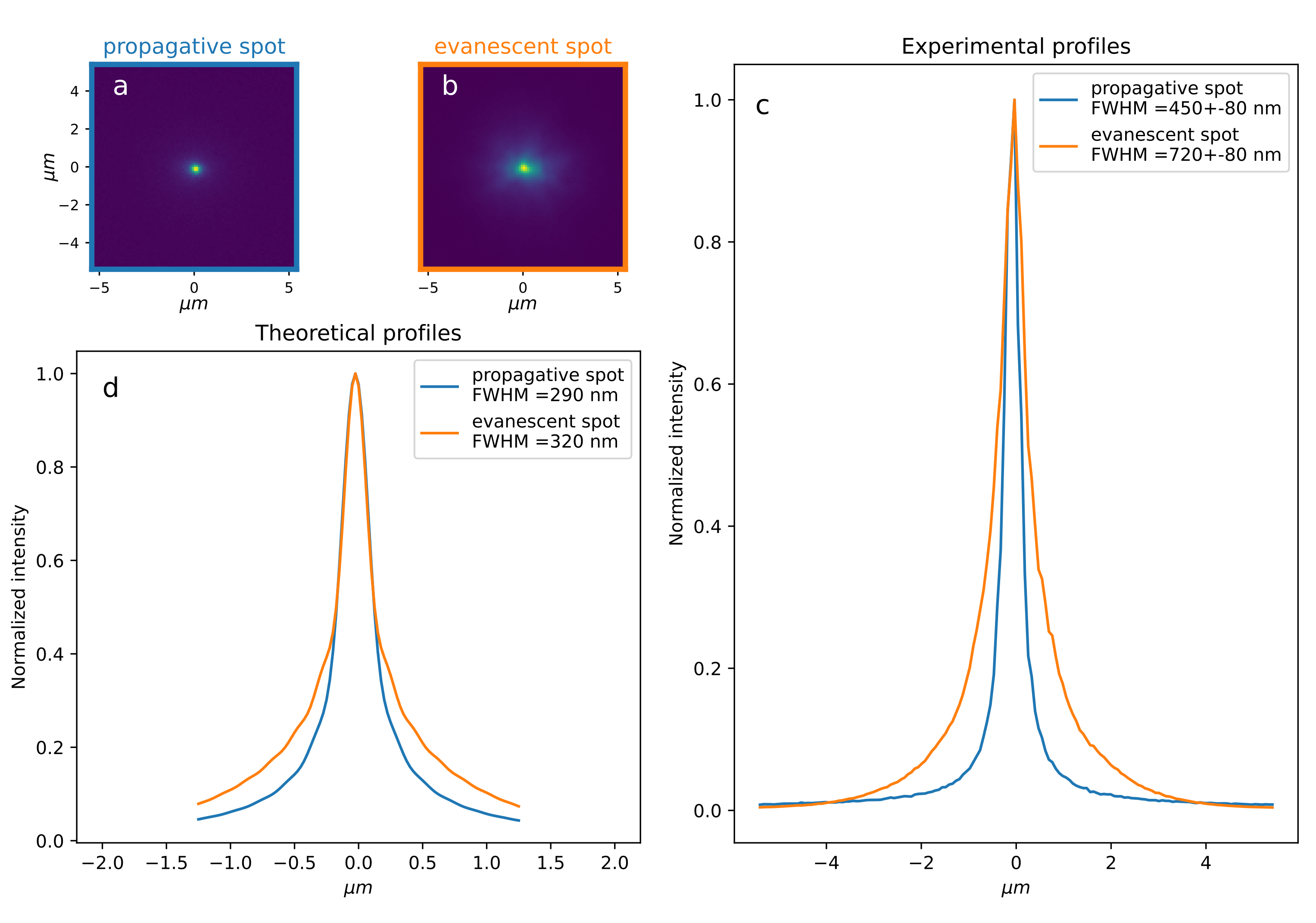}
    \caption{Lateral width characterisation. \textit{(a,b)} Fluorescence images, in the ($x,y$) plane, of a propagative spot by illuminating the UAR (\textit{blue}) and an evanescent spot made by illuminating the SAR (\textit{orange}) at the interface between the coverslip and a dye solution. \textit{(c)} Normalized radial profiles extracted from these images. \textit{(d)} Theoretical profiles computed with a scalar model.}
    \label{lateral_res}
\end{figure*}

Fig.~\ref{lateral_res}(a,b) depict fluorescent images of the spot in the two cases, while Fig.~\ref{lateral_res}(c) shows the profiles computed from the images by radial averaging around the spot center.  In the case of an evanescent focus (SAR), the fluorescent spot is wider, compared to the case of a propagative focus (UAR). The full widths at half maximum (FWHM) are respectively 0.72~\textmu m and 0.45~\textmu m. Moreover, a blurry region surrounds the central spot, which is visible on the evanescent spot profile through the presence of side wings.

We have confronted our experimental observations to theoretical predictions for a scalar optical field, in both the UAR and SAR cases. More precisely, the excitation beam has been described as a spherical wave truncated by the SAR or UAR pupil. Using the angular spectrum representation~\cite{novotny2012principles}, the incoming wave has been propagated up to the interface between the two media, then in the sample itself (aqueous medium). For each given plane in the sample, the convolution of the excitation intensity distribution with the detected point spread function (also calculated using angular spectrum representation) is computed to provide the corresponding observed point spread function. Finally, the contributions of planes up to 5~\textmu m in depth were added up to account for the sample thickness. The theoretical calculations reproduce the side wings observed on the experimental evanescent profile: indeed, although an annular aperture leads to a focus resembling a Bessel beam with numerous outer rings~\cite{Ramsay2008} the convolution with the detection point spread function smoothed out the rings, leading to the observed side wings (Fig.~\ref{lateral_res}(d)).

However, the size of the experimental spot is not in agreement with the theoretical prediction. The computed spots have FHMH of 0.32~\textmu m and 0.29~\textmu m, for SAR and UAR illuminations respectively, which is almost twice smaller than the experimental values (Fig.~\ref{lateral_res}(d)). We attribute this difference to remaining aberrations that have not been successfully corrected in our experiment, possibly in high order modes. The difference between the experimental and the theoretical spot is larger for SAR illumination. Indeed, numerical simulations have shown that, when light passes through the SAR region, the spot shape is more impacted by aberrations than when the beam fills the UAR disk.

In previous work~\cite{Dorn2003},  radially polarized beams have been shown to lead to smaller spot sizes. Hence, we added a radial polarization converter (S-waveplate, RPC-488-10-1007) at the back entrance of the objective, in order to create a radially polarized beam in the BFP. However, we did not notice any measurable difference on the size of the fluorescent spot. This further suggests that, in our case, the spot size is limited by the remaining optical aberrations, thus hiding the effect of polarization.

\subsection*{Penetration depth}

\begin{figure}[ht!]
    \centering

    \includegraphics[width=1\linewidth]{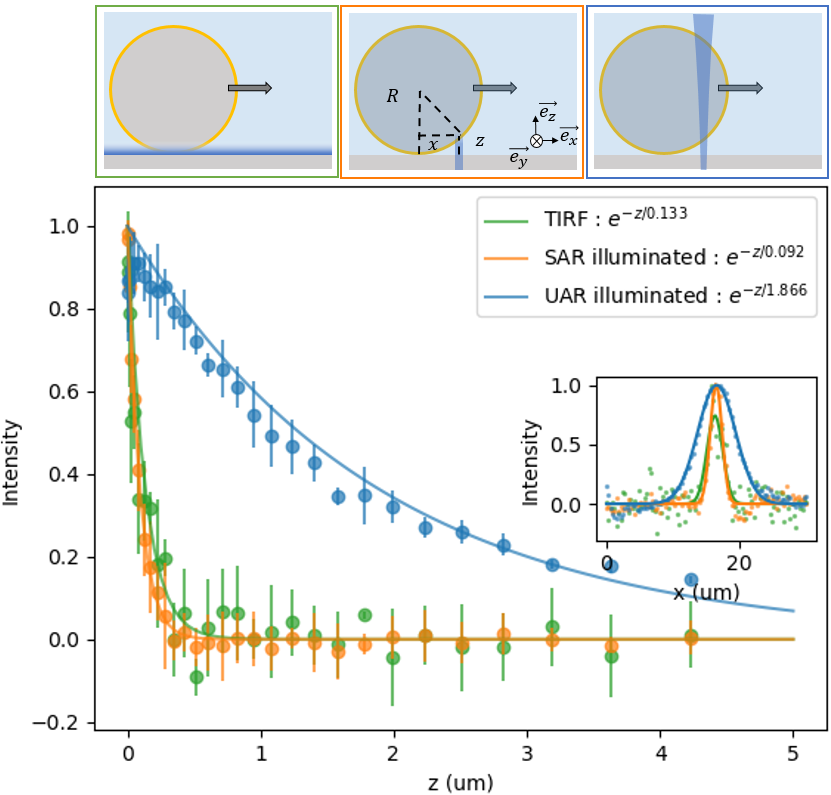}
    \caption{Penetration depth measurement. A bead of radius, $R \approx 5 $ \textmu m, covered with fluorescent molecules, is scanned along $x$ and 3 configurations are studied \textit{(top drawings)}: standard TIRF \textit{(left green box)}, evanescent spot with SAR illumination \textit{(centre orange box)}, propagative spot with UAR \textit{(right blue box)}. We measured the intensity detected by the pixel located at the spot centre as a function of $x$ \textit{(inset)}. The background level, when the bead is not above the detected position, is set to 0 and the peaks, corresponding to passing bead, are normalized in order to compare the different configurations. On the main plot, the intensity is represented as a function of the vertical distance $z$, after converting $x$ to $z$ values. Data over N=5 beads have been collected, and the error bars correspond to their standard deviation.}
    \label{axial_res}
\end{figure}

Multiple configurations have been proposed~\cite{Oheim2019} to measure the penetration depth of a standard TIRF, but many of them are not suited to the situation of a point-like evanescent wave. Here, we adapted the method described in Ref.~\cite{Mattheyses2006} that employs large beads coated by fluorescent molecules. We used $10$~\textmu m polystyrene beads (TPX-100-5, Spherotech), coated with streptavidin-functionalized AlexaFluor-488 (Life Technologies, S32354), prepared according to the protocol described in~\cite{Cabriel2018}, diluted in water and deposited on a glass coverslip. 

The measurement consists in scanning one bead over the evanescent spot along the $x$ axis, as shown on Fig.~\ref{axial_res}. For comparison, two other configurations have been studied: a classical standard TIRF achieved by illuminating a single focus in the SAR of the objective BFP and a propagative spot achieved by illuminating the UAR. In all cases,  an image was recorded by the camera for each bead position, and we extracted the fluorescence intensity of the pixel corresponding to the spot center (or, in the case of standard TIRF, an arbitrary pixel) . The resulting intensity as a function of $x$ is shown on the inset of Fig.~\ref{axial_res}.

Assuming that the bead is touching the coverslip, the lateral position $x$ can be converted to the vertical distance between the glass/water interface and the fluorescent layer, $z$, by the relation: $z = R - \sqrt{R^2 - x^2}$ with $R$ the radius of the bead and $x = 0$ at the contact point. Hence, the fluorescence intensity can be plotted as a function of depth $z$, as depicted in Fig.~\ref{axial_res}. Here we assumed that each measured intensity is unambiguously associated to a single $z$ value, which is not strictly correct since the area on the sphere illuminated by the spot is not point-like. However, since $z$ varies very slowly with $x$ in the range of interest, this effect was considered to be negligible.

The data of Fig.~\ref{axial_res} show a fast decay of the fluorescence intensity as a function of depth, both in the case of evanescent spot (SAR) and standard TIRF. By fitting these curves to an exponential decay, we found characteristic lengths of $\delta_{\rm{TIRF}} = 100 \pm 15 \ \rm{nm}$ and $\delta_{\rm{SAR}} = 93 \pm 3 \ \rm{nm}$. In the case of a propagative spot (UAR), the detected intensity is also found to decrease with $z$, due to the defocus that causes the spot to spread over a larger area and affects both excitation and detection paths. The curve obtained with a propagative spot enables to evaluate this effect. Still using an exponential fit, we found a decay length of $\delta_{\rm{UAR}} = 1675 \pm 50 \ \rm{nm}$, which is an order of magnitude larger than the decay found for an evanescent spot or standard TIRF. Hence, the effect of defocus is negligible on the measured evanescent depth. 

Finally, these data confirm that the spot generated by SAR illumination is truly evanescent, with a penetration depth of around 100~nm comparable to standard TIRF.

\section*{Optogenetic activation in living cells}

\begin{figure*}[ht!]
    \centering
    \includegraphics[width=0.7\linewidth]{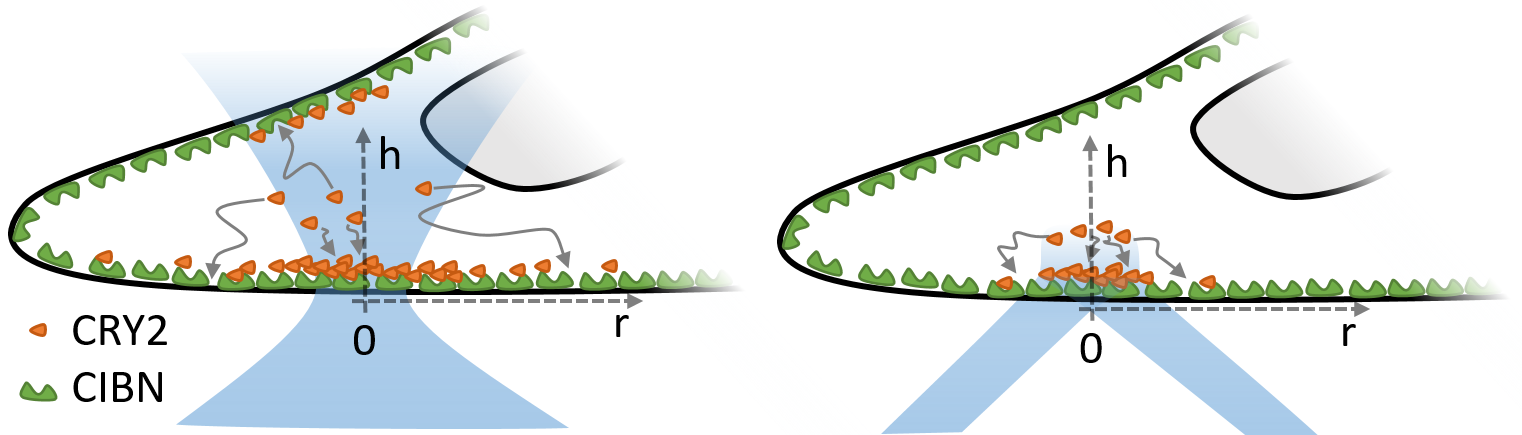}
    \includegraphics[width=0.7\linewidth]{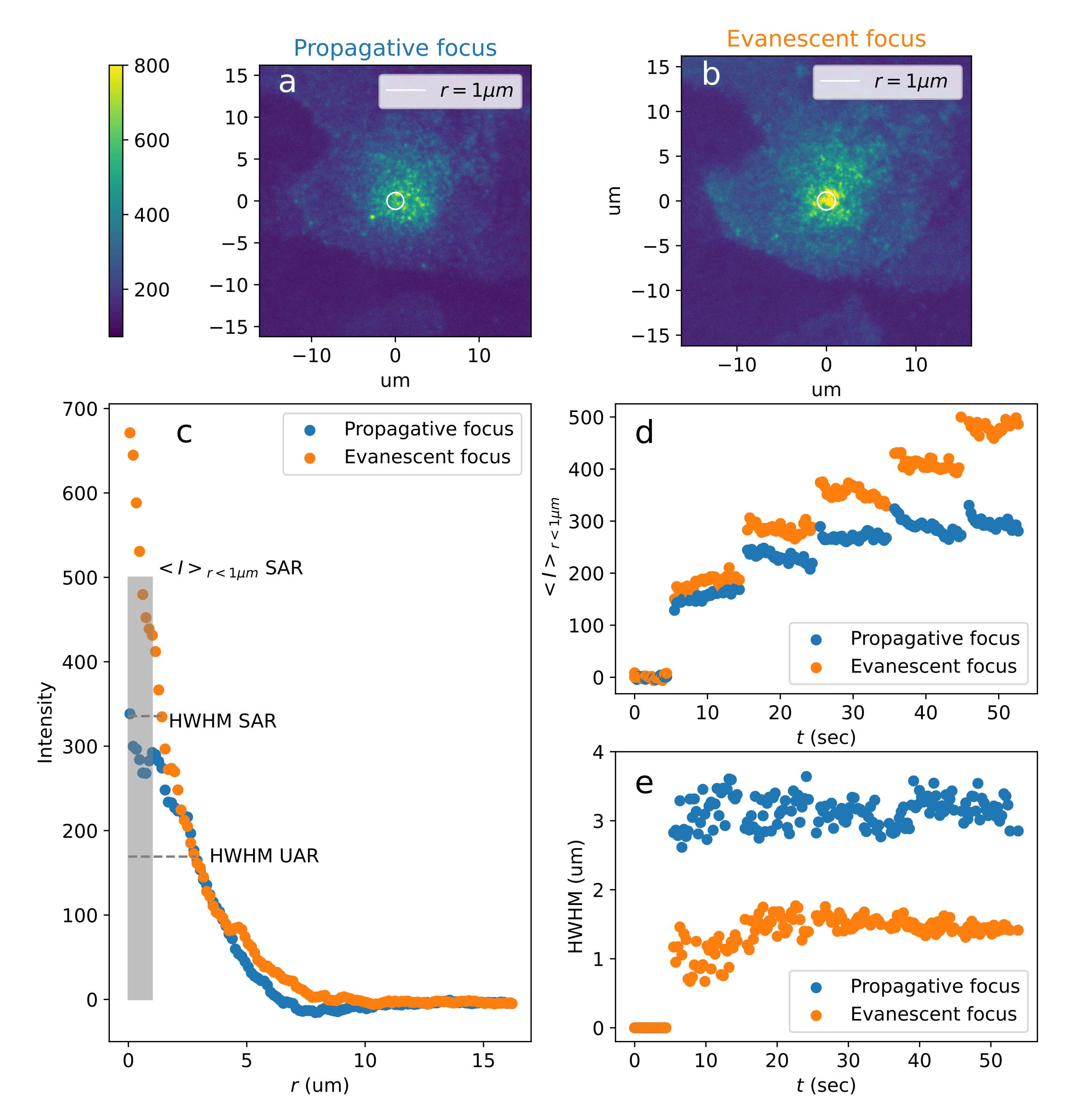}
    \caption{Optogenetic experiment. \textit{Top:} Schematic view of the recruitment of CRY2 at the membrane under a propagative and an evanescent activation. Standard TIRF images of the cell membrane after \textit{(a)} a series of evanescent and \textit{(b)} a series of propagative pulses. \textit{(c)} Examples of intensity profile obtained by radial averaging. \textit{(d)} Average intensity in a 1~\textmu m-radius disk as a function of time. \textit{(e)} Half width at half maximum, estimated from the profiles (dashed lines in c) as a function of time. \textit{Blue:} propagative focus. \textit{Orange:} evanescent focus.}
    \label{cell_SAR_vs_UAR}
\end{figure*}

We investigated the potential of an evanescent spot to improve the spatial resolution of optogenetic perturbations, using the popular CRY2/CIBN hetero-oligomerizer system. Upon blue illumination, activated CRY2 (fused to the fluorescent protein mCherry) in the cytosol changed its configuration revealing binding sites to its partner CIBN which is tethered at the membrane through a CAAX motif tagged with GFP (Green Fluorescent Protein). We studied the local recruitment of CRY2-mCherry at the membrane by TIRF imaging (at 561 nm) after stimulation with either a propagative or an evanescent focus (Fig.~\ref{cell_SAR_vs_UAR}).

Experiments were performed on epithelial Madin-Darby canine kidney (MDCK) cells stably expressing CRY2-mCherry and CIBN-CAAX using a lentiviral transfection~\cite{Kerjouan2021}. The cell line was cultivated at 37°C and 5\% CO$_2$ in DMEM (Dulbecco’s Modified Eagle Medium) high glucose (4.5 g/l) with glutamax medium (PAA Laboratories) supplemented with 10\% fetal bovine serum (GE Healthcare) and penicillin and streptomycin 1\% (v/v) (PAA). For live imaging experiments, MDCK cells were plated on 2-well Labtek coverslips and incubated in CO$_2$-independent medium (Thermofisher scientific, ref:18045088) complemented with 10\% fetal bovine serum (GE Healthcare).

Since CRY2 is labelled by mCherry, the 561~nm standard TIRF path is used to monitor CRY2 recruitment at the basal plasma membrane of the cell, while the 488~nm laser is used for CRY2 photoactivation. During the experiment, a cell is first illuminated  by a series of pulses in propagative mode (UAR illuminated, 5 pulses of 100~ms, each spaced out by 10~s). Then, we wait a 10~min-period in the dark to allow a full dissociation of the CRY2-CIBN hetero-oligomers since the dissociation time of CRY2-CIBN is 4.5~min~\cite{Kennedy2010} and the same cell is submitted to a series of pulses in evanescent mode (SAR illuminated) with the same temporal parameters.

To set the laser power, we chose a cell different from the one used in the final experiment and iteratively increased the power (while in propagative mode) until a recruitment of CRY2 proteins is observed. In order to have comparable quantities of photons between both modes of activation, evanescent versus propagative, we took into account the difference between the SAR and UAR areas on the DMD. Thus, we adjusted the initial laser power so that the amount of light that reaches the sample stays unchanged in both evanescent and propagative modes (P=4~\textmu W). 

The cell is continuously imaged under 561~nm TIRF excitation, starting approximately 10~s before the first pulse. The images recorded before activation have been averaged and subtracted to each image, prior to extracting the profiles shown in Fig.~\ref{cell_SAR_vs_UAR}c. Since CRY2 forms oligomers upon illumination as seen in Fig.~\ref{cell_SAR_vs_UAR}(a,b) and in previous works~\cite{Bugaj2013}, we performed radial averaging to obtain a smooth intensity profile at each time-point (Fig.~\ref{cell_SAR_vs_UAR}c). Then the intensity within a 1~\textmu m-radius disk around the focus (Fig.~\ref{cell_SAR_vs_UAR}d) and the half width at half maximum (HWHM) (Fig.~\ref{cell_SAR_vs_UAR}e) have been extracted from each profile.

First, our data show that an evanescent spot induces a more localized membrane recruitment, compared to a propagative spot, since we observe a significant reduction of the HWHM of CRY2 membrane distribution after activation (Fig.~\ref{cell_SAR_vs_UAR}e): $\rm{HWHM}_{\rm{SAR}} = 1.65 \pm 0.15 $~\textmu m versus $\rm{HWHM}_{\rm{UAR}} = 3.0 \pm 0.15$~\textmu m. 
This narrowing of the CRY2 recruitment region after evanescent pulses shows that numerous activated CRY2 bind immediately to the membrane and do not diffuse in the cytoplasm. On the contrary, the recruitment after a propagative pulse seems to be strongly affected by cytosolic diffusion, as CRY2 is recruited over a large region. 
Since a propagative focus generates a conical volume of activated CRY2 molecules throughout the thickness of the cell, the molecules diffuse in the cytoplasm before binding on the membrane observed in TIRF. The typical lateral distance they would travel (from the focus) until they bind to the basal membrane is related to their average activation depth. This depth is drastically reduced in the case of an evanescent focus, resulting in a smaller recruitment region. However, diffusion may still occur before the binding event; this would explain why the HWHM of CRY2 distribution is larger than the size of the illumination spot (as measured in Section~\ref{lateral_size}) which is sub-micron. Previous work~\cite{Valon2015} has pointed to the role of 2D diffusion of CRY2-CIBN dimers in the membrane once they have formed, as a factor that limits the size of the activated region. In our case, however, the width of CRY2 membrane distribution does not increase over tens of seconds, suggesting that membrane diffusion is much slower than the value measured previously ($\approx$ 0.1~\textmu$\rm{m}^2$/s~\cite{Valon2015}).

On top of an improved spatial resolution, Fig.~\ref{cell_SAR_vs_UAR}d shows that an evanescent spot can induce a higher intensity, hence more recruited proteins at the peak, compared to a propagative spot. Interestingly, the first pulse induces the same number of recruited CRY2 (the height of the first step on the two curves is similar), which confirms that we have correctly adjusted the illumination power to be equal in both configurations. But, while the number of proteins recruited after each pulse remains mostly unchanged in the case of an evanescent focus, it reduces in the case of propagative activation. This reduction can be attributed to the progressive depletion of cytosolic CRY2, as already reported before~\cite{Valon2015}: since more proteins are bound to the membrane, the cytosolic concentration decreases and fewer proteins can be activated at each pulse. This depletion of CRY2 in the cytoplasm is more pronounced in the case of a propagative spot, as the latter illuminates a large volume of the cytoplasm and could induce CRY2 recruitment in other parts of the cell (such as the apical membrane) that are not visible in our images, which contributes to decreasing the cytosolic concentration. The evanescent spot illuminates only a very small volume and maximizes the recruitment at the point of interest, so that CRY2 depletion in the cytoplasm is almost negligible and each pulse shows the same efficiency in terms of recruitment. In conclusion, one can reach a much higher density of CRY2 at the desired location using an evanescent focus.

Our preliminary results on living cells show that evanescent focus illumination can significantly improve both the spatial resolution and the recruitment efficiency, paving the road to sub-micrometric and organelle-resolution optogenetic approaches. To better understand these observations and improve the spatio-temporal control of optogenetic activation, further work is ongoing to study the role of various molecular processes in cells (cytosolic and membrane diffusion, dissociation, etc.)

\section*{Conclusion}

In this work, we have presented a method to create an evanescent spot, confined in the three dimensions of space, which can be moved across the field of view. This evanescent focus is produced by sending light in a  ring at the periphery (corresponding to the super-critical region) of the BFP of a high numerical aperture objective, using a DMD. Compared to a liquid crystal spatial light modulator or a deformable mirror which also allow wavefront shaping, DMDs exhibit lower light efficiency, since they only display binary patterns (leading to symmetric diffraction orders around the 0th order).  However they have the advantages of a fast switching speed and a high number of uncoupled degrees of freedom, which allows the DMD in our setup to serve as annular mask, beam scanner and wavefront corrector.
Thanks to the high frame rate of the DMD (10 kHz), the evanescent focus can be rapidly scanned in order to  create an evanescent pattern. 
We measured both the lateral size and the penetration depth of the evanescent focus. Although the diameter of the evanescent spot is sub-micron ($\rm{FHMH} = 780$~nm), we found that it is very sensitive to aberrations and remains significantly larger than a propagative focus obtained in similar conditions. The penetration depth of this evanescent spot, however, was found to be similar to that of standard TIRF illumination ($\sim$100~nm).
We also explored the interest of evanescent spot illumination for optogenetics by activating the CRY2-CIBN hetero-oligomerizer system. When induced by an evanescent compared to a progagative focus, the size of the recruitment region is reduced. Moreover, a higher density of recruited proteins can be reached at the position of interest under evanescent illumination, since the amount of proteins that bind elsewhere in the cell is reduced. Evanescent photoactivation could help understanding biological processes at the vicinity of the basal membrane, such as cell adhesion and migration.
Hence, using an evanescent focus in future seems highly promising to go from cellular to subcellular optogenetic experiments.

\textbf{Funding} Agence Nationale de la Recherche (ANR-20-CE42-0018); Association de Recherche contre le Cancer (ARC 2021 PJA3)

\textbf{Acknowledgements} The authors acknowledge Philippe Moreau for technical support on the setup, Jean Revilloud for advice on cell culture, Christiane Oddou for molecular biology, Cristina Torres and Lucia Campos-Perello for preparing the cells used in this work.

\end{document}